\documentclass{jetpl} %for JETP pdf file%
\twocolumn  %for JETP pdf file%
\lat

\title{Polyelliptic  coordinates for solving the Schr\"{o}dinger  and  Helmholtz equations 
}

\rtitle{Polyelliptic  coordinates\ldots}

\sodtitle{Polyelliptic coordinates for solving the Schr\"{o}dinger  and  Helmholtz equations. I.
}

\author{G.\,V.\,Kovalev \/\thanks
{e-mail: kovalevgennady@qwest.net}}

\rauthor{G.\,V.\,Kovalev}

\sodauthor{Kovalev}

\address{North Saint Paul,  MN 55109, USA \\[5mm]}

\dates{19 February 2013}

\abstract{Several local elliptic coordinates are used to build a new polyeliptic coordinate system  which is orthogonal and admits the separation of variables. Such coordinate systems can give the exact solutions of some unsolved problems in quantum mechanics and diffraction theory. 
}  

\begin{document}

\maketitle

\textbf{I.Introduction.} The most powerful method for solving  problems of quantum mechanics, diffraction theory and mathematical physics, in general, is the method of separation of variables.  More than 60 years ago Robertson and Eisenhart \cite{RobEis, MoonSpencer:1988,MorseFeshbach:1953} have shown that only 11 different coordinate systems allow the separation of the Schr\"{o}dinger (SE) or  Helmholtz (HE) equations in 3D space. The  ellipsoidal system is the most general, the remaining 10 are the special cases of ellipsoidal. In particular, 3D Cartesian coordinates can be obtained from ellipsoidal when $\{f_1,f_2\} \rightarrow \infty $  ($f$ is semi-focal distance) and the traditional spherical coordinates are received when $ \{f_1,f_2\} \rightarrow 0$. The theory of separation of variables is well developed\cite{Miller:1976,Kalins:1986} and has many advances  since the time of publications of \cite{RobEis}, but the ellipsoidal coordinates are still intriguing and many properties of ellipsoidal wave functions are unknown \cite{Fedoryuk:1989, PDG:2004,Slepian:1983}.

The purpose of this work is to show that the conclusion of Robertson and  Eisenhart \cite{RobEis} is restricted to a single ellipsoidal coordinate system. But if we use  several local ellipsoidal coordinate systems and combine them together  by special rules described below, we can get  a very large number of coordinate systems which  are orthogonal  and admit the separation of variables. In fact, even in 2D space the number of new coordinate systems is infinite because each new  system  is related to one specific  convex polygon (or cone, polyhedron, finite cylinder, etc. for 3D case).  The 3 examples for 2D space are built here  and the separated ordinary differential equations for  HE  in new 2D coordinate system are derived in general form. 
On the reason which is clear from the context, these coordinates systems can be called  'polyelliptic' for 2D  and 'polyellipsoidal', 'polyspheoidal', etc. for 3D spaces.

\begin{figure}
	\centering
		\includegraphics[width=0.50\textwidth]{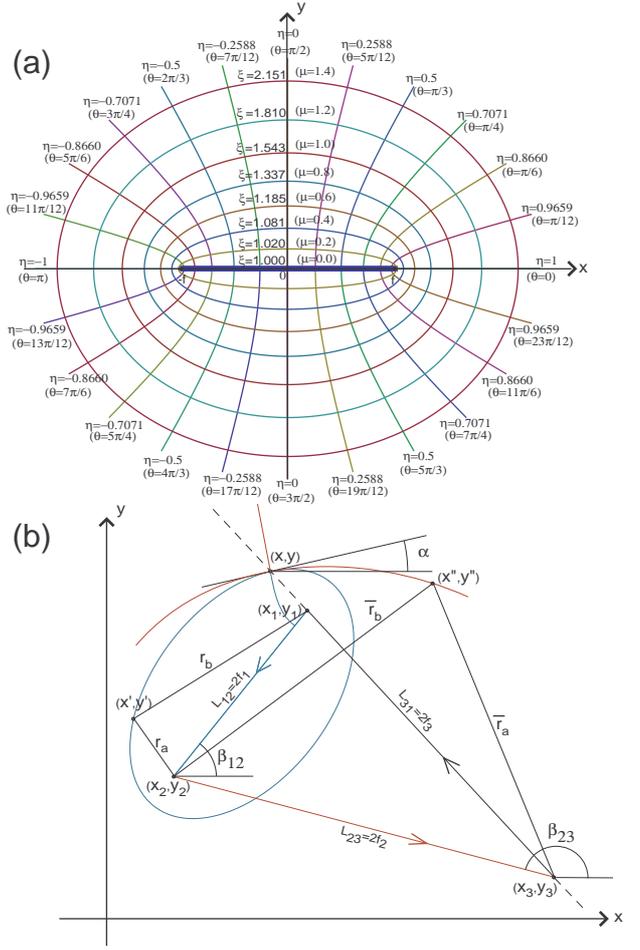}
	\caption{Fig.1. (a) The elliptic coordinates ${\xi, \eta}$: $x=f \xi \eta,\; y=f\sqrt{(\xi^2-1)(1-\eta^2)}$. Upper and lower semi-hyperbolas are different coordinate lines, therefore instead ${\xi, \eta}$ one should use  $\mu, \theta$:  $\xi=\cosh \mu$, $\eta=\cos\theta$. (b)Two elliptic coordinate systems with focal distances $L_{12}$, $L_{23}$ and common focal point $(x_2,y_2)$  have the same tangent derivative $(\frac{d y(x)}{d x})_{12}=(\frac{d y(x)}{d x})_{23}=\tan{\alpha}$ to their ellipses (blue for $L_{12}$ and red for $L_{23}$)  on dashed straight line coming out from the point $(x_1,y_1)$. The same is true for semi-hyperbolas.}
	\label{fig:EllipticCoords}
\end{figure}

\begin{figure}[htbp]
	\centering		
	\includegraphics[width=0.50\textwidth]{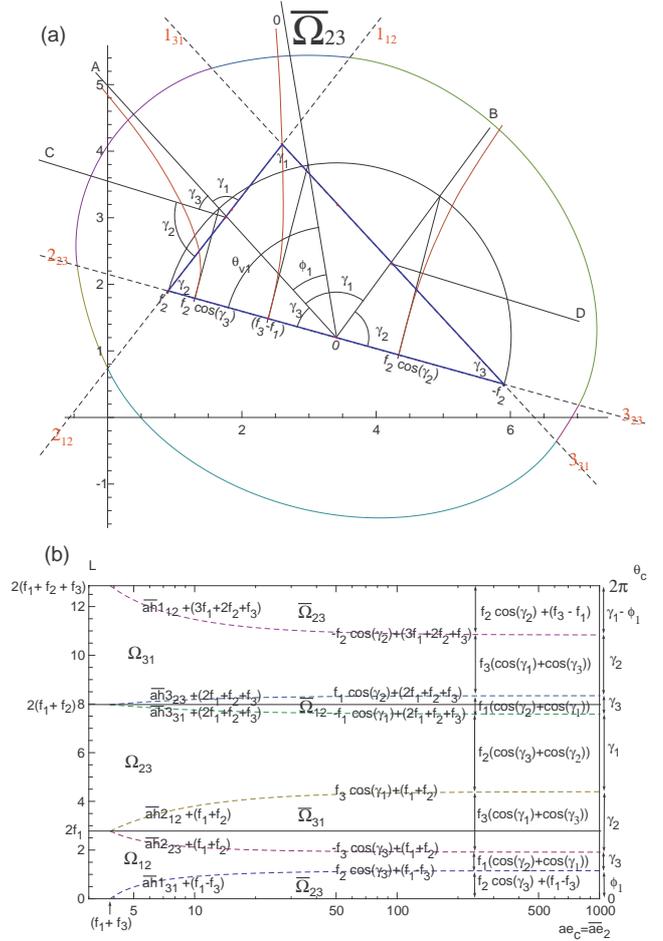}
	\caption{Fig.2. (a)The triangle 'polyelliptic' coordinates $\mu_c, \theta_c$ are built from 3 local elliptic coordinates: $L_{12}=2.78$,  $L_{23}=5.19$,  $L_{31}=4.88$; $\gamma_1=80.2^{\circ}$,  $\gamma_2=67.95^{\circ}$, $\gamma_3=31.85^{\circ}$. (b) All  horizontal  straight lines present the hyperbolic or hyperbolic-like coordinates; the curved dashed lines present the the straight lines $1_{31},2_{23},...$. }
	\label{fig:CommonCoords+EllipticEgg1}
\end{figure}

\begin{figure}[htbp]
	\centering
\includegraphics[width=0.50\textwidth]{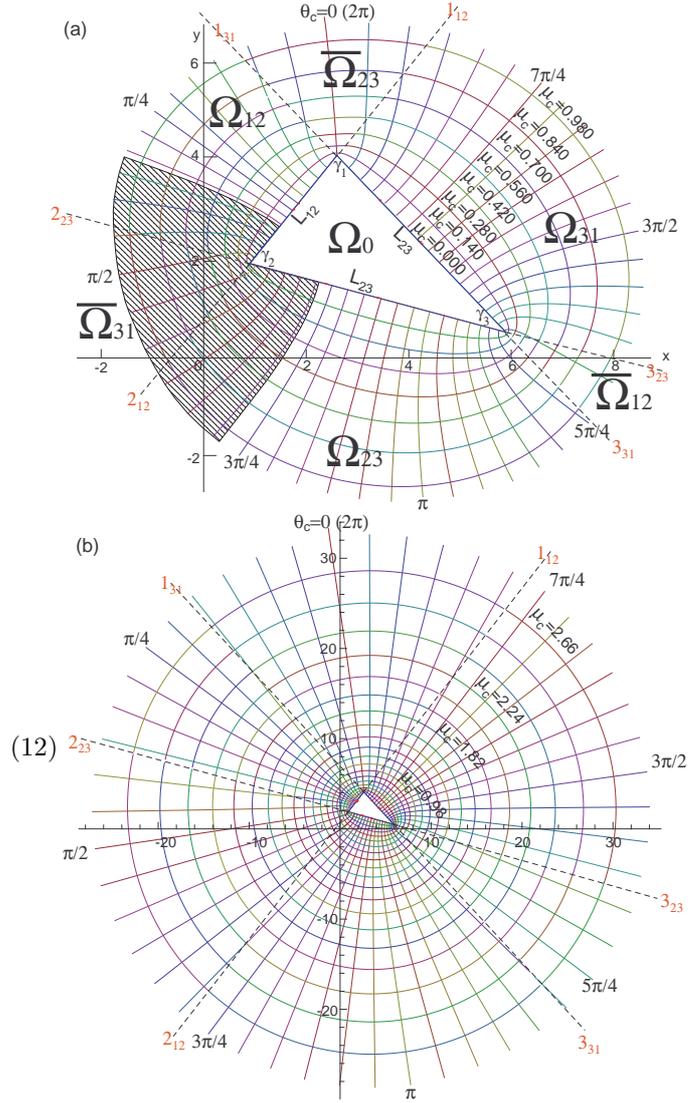}
	\caption{Fig.3  (a)  The final structure of arbitrary triangle  'polyelliptic' coordinates $\mu_c, \theta_c$  with: $L_{12}=2.78$,  $L_{23}=5.19$,  $L_{31}=4.88$; $\gamma_1=80.2^{\circ}$,  $\gamma_2=67.95^{\circ}$, $\gamma_3=31.85^{\circ}$. Shade at one corner shows the area of hyperbola-like coordinates.  (b) The same coordinate system for large  $\mu_c$  becomes similar to the polar coordinates.}
	\label{fig:TriangleCoords_Small-Large}
\end{figure}

\begin{figure}
	\centering
\includegraphics[width=0.50\textwidth]{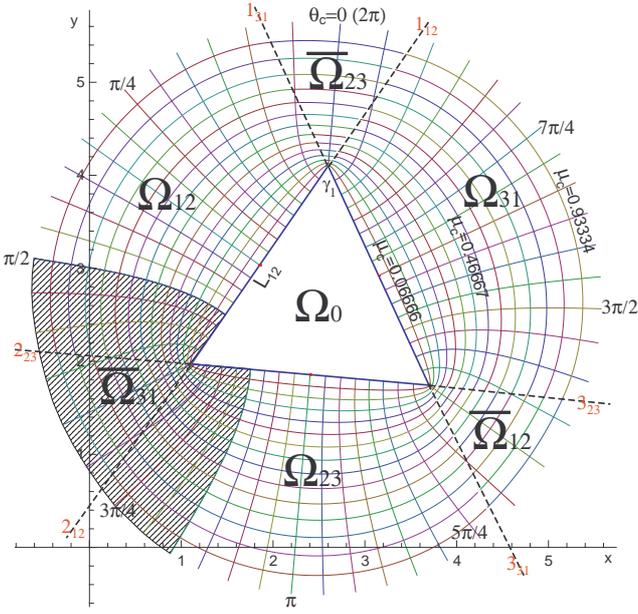}
	\caption{Fig.4. The equilateral triangle  'polyelliptic' coordinates $\mu_c, \theta_c$ with $f_1=f_2=f_3=1.3$. Shade at one corner shows the area of hyperbola-like coordinates.}
	\label{fig:EquilateralEllipticTriangle_Small}
\end{figure}

To illustrate the construction method, we consider 2D case where only 4 orthogonal coordinate system having the exact solutions of HE are known \cite{Miller:1976} and where elliptic coordinates, Fig.1(a), are the most general. The higher dimensional cases can be constructed using the similar rules and ideas.   To build the new coordinates, we take three or more local elliptic coordinates  with  semi-focal distances $f_1, f_2, f_3,...$ and put them on one plane letting the pair of focal points coincide in such a way that the interfocal line segments, denoted as $L_{ij}$:   $L_{12}=2 f_1$, $L_{23}=2 f_2,...$, constitute an arbitrary convex polygon (see, e.g., triangle in Fig. \ref{fig:EllipticCoords}(b)). Then we can observe the following properties of such construction: 

\begin{itemize}
\item On the dashed lines $N_{ij}$  in Fig.\ref{fig:EllipticCoords}(b),\ref{fig:CommonCoords+EllipticEgg1}(a),\ref{fig:TriangleCoords_Small-Large},\ref{fig:EquilateralEllipticTriangle_Small} ($N$ is  the numeric label of the vertex,  $\{i,j\}$ are indices of a side of the polygon) the coordinate lines of 2 elliptic systems, i.e. ellipses and semi-hyperbolas,  have the same tangent direction. Therefore, the coordinate lines at crossing the dashed lines preserve their directions and can be switched from one coordinate system to another. \footnote{The short proof of this statement is in Fig.\ref{fig:EllipticCoords}(b), detailed proof will be published elsewhere.}

\item The dashed lines and sides of n-polygon separate the XY-plane  into $2n$ regions: $\Omega_{12}$, $\overline{\Omega}_{12}$, $\Omega_{31}$, $\overline{\Omega}_{31}$,... . Each region has a local elliptic coordinate system generated by one interfocal line, e.g. the region $\Omega_{12}$ is generated by $L_{12}$, the region $\overline{\Omega}_{23}$ generated by left side of $L_{23}$\footnote{We also introduce the 'polarization' of a focal line segment $L_{ij}$; the positive direction is from focus $f$ to negative focus $-f$ and on the right side of $L_{ij}$  the local elliptic coordinate $\mu_{i}$ has the range $0 \leq \theta_{i}\leq \pi$, on the left side the range is $\pi \leq \theta_{i}\leq 2\pi$. For short, the left side is denoted by bar always.  In such conventions, the positive path around a convex polygon is the counterclockwise, Fig. \ref{fig:EllipticCoords}(a), and region $\overline{\Omega}_{23}$ is generated by left side of the interfocal line $L_{23}$.}, etc..   
\item The coordinate lines of new common system are smooth everywhere outside the  perimeter of the polygon and it is possible to introduce a common 'polyelliptic' coordinates $\mu_c$, $ \theta_c$ for this region including the perimeter.
\item All congruent polygons have the similar type of 'polyelliptic' coordinate system. The HE and SE with potential function having St\"{a}ckel form for these system gives a new type of orthogonal eigenfunctions and eigenvalues similar to Mathieu's types. Noncongruent  polygons give the different  'polyelliptic' coordinates and different types of orthogonal eigenfunctions and eigenvalues.  
\item There is a 'protected area' $\Omega_0$ (interior of the convex polygon), which is not covered by this method. However, in many cases the $\Omega_0$  can be covered by another orthogonal coordinates (e.g., Cartesian, as in Fig.\ref{fig:SquareWell_SquareShape_3a}). In some cases (diffraction, QM), this coverage is not necessary because the sides of polygon can satisfy Dirichlet's conditions.
\item When one semifocal distance $f_i \rightarrow 0$, the 'polyelliptic' coordinates generated by n-polygon  is topologically transformed to  another 'polyelliptic' system generated by (n-1)-polygon. In particular, when $f_3 \rightarrow 0$ the triangular 'polyelliptic' coordinates are transformed to an ordinary elliptic coordinate system.  Hence, a single elliptic coordinate system Fig.\ref{fig:EllipticCoords}(a) is a particular case of the general 'polyelliptic' coordinates Fig.\ref{fig:TriangleCoords_Small-Large}(a).  
\end{itemize}
\textbf{II. Triangle Elliptic Coordinates and other.} The local elliptic coordinate system $(\mu_i,\theta_i)$ in region  $\Omega_{ij}$ or $\overline{\Omega}_{ij}$ for arbitrary orientation and displacement of the focal line $L_{ij}$ is defined as 
\begin{eqnarray}
x&=A_i \cos{\beta_{ij}} -B_i\sin{\beta_{ij}}+ \frac{x_i+x_j}{2} , \nonumber \\
y& =A_i \sin{\beta_{ij}} +B_i \cos{\beta_{ij}}+\frac{y_i+y_j}{2},
\label{eq1}
\end{eqnarray}
where $A_i,B_i$ are usual elliptic transformation functions:
\begin{eqnarray}
A_i&=f_i \cosh{\mu_{i}} \cos{\theta_{i}}, \;\;\;
B_i& =f_i \sinh{\mu_{i}} \sin{\theta_{i}},
\label{eq1a}
\end{eqnarray}
and $\beta_{ij}$ is the angle between negative direction of focal distance $L_{ij}$ and positive direction of axis $x$, Fig. \ref{fig:EllipticCoords}(b). When two points $(x',y')$,  $(x'',y'')$ coincide on the dashed semi-infinite line, the relations between two major elliptic semi-axes: $ae=(r_a+r_b)/2$, $\overline{ae}=(\overline{r}_a+\overline{r}_b)/2$, and two major hyperbolic semi-axes: $ah=(r_a-r_b)/2$,  $\overline{ah}=(\overline{r}_a-\overline{r}_b)/2$,  are $\overline{ae}-ae=f$ and $\overline{ah}+ah=\pm f$. These relations in triangle  'polyelliptic' coordinates for all  major elliptic and hyperbolic semi-axes are (1-st column denotes the dashed lines in Fig. \ref{fig:TriangleCoords_Small-Large}(a)):
\begin{eqnarray}
1_{31}:&\;\; \overline{ae}_2-ae_1=f_3, \;\;\; \overline{ah}1_{31}+ah1_{31}=f_3;\nonumber \\  
2_{23}:&\;\; \overline{ae}_3-ae_1=f_2, \;\;\; \overline{ah}2_{23}+ah2_{23}=-f_2;\nonumber \\  
2_{12}:&\;\; \overline{ae}_3-ae_2=f_1, \;\;\; \overline{ah}2_{12}+ah2_{12}=f_1;\nonumber \\  
3_{31}:&\;\; \overline{ae}_1-ae_2=f_3, \;\;\; \overline{ah}3_{31}+ah3_{31}=-f_3;\nonumber \\  
3_{23}:&\;\; \overline{ae}_1-ae_3=f_2, \;\;\; \overline{ah}3_{23}+ah3_{23}=f_2;\nonumber \\  
1_{12}:&\;\; \overline{ae}_2-ae_3=f_1, \;\;\; \overline{ah}1_{12}+ah1_{12}=-f_1. 
\label{eq2}
\end{eqnarray}

The similar relations hold for any polyelliptic system generated by a convex n-poligon and allow to introduce the common elliptic coordinates. Indeed, all major elliptic semi-axes \eqref{eq2}  in all regions can be expressed through just one, say $\overline{ae}_2$:
\begin{eqnarray}
ae_1=\overline{ae}_2-f_3, \;\;\; \overline{ae}_3=\overline{ae}_2-f_3+f_2,\nonumber \\  
ae_2=\overline{ae}_2-f_3+f_2-f_1, \;\;\; \overline{ae}_1=\overline{ae}_2+f_2-f_1, \nonumber \\  
ae_3=\overline{ae}_2-f_1, 
\label{eq3}
\end{eqnarray}
so the value $\overline{ae}_2$ can be considered as a common 'radial' coordinate $ae_c$.
We note, however, that the range  of $ae_c=\overline{ae}_2$  is  from  $f_1+ f_3$ to $\infty$ and all others $ae$'s  are positive. From $ae_c$, we can deduce  another 'radial' common coordinate $\mu_c$ with 'universal' range:     
\begin{eqnarray}
\mu_c=\text{arccosh}[\frac{ae_c}{f_1+f_3}], \;\;\; 0\leq \mu_c \leq \infty. 
\label{eq4}
\end{eqnarray}
We can take any other elliptic semi-axes $ae$ as a common 'radial' coordinate using \eqref{eq2}. The respective transition from $ae$  to  $\mu_c$  will give the similar  'radial' common coordinate $\mu_c$  with the same range $ 0\leq \mu_c \leq \infty$.
 
The angular common coordinate $\theta_c$ is also defined from \eqref{eq2}, but its definition  is more complicated. For our triangle, the hyperbolas lying in the ranges of local major hyperbolic semi-axes, 
\begin{eqnarray}
-&\cos \gamma_2 \leq \frac{ah_1}f_1 \leq  \cos \gamma_1, \;\;
-\cos \gamma_3 \leq \frac{ah_2}f_2 \leq  \cos \gamma_2, \;\; \nonumber \\
&-\cos \gamma_1 \leq \frac{ah_3}f_3 \leq  \cos \gamma_3,  
\label{eq5}
\end{eqnarray}
remains the same and incorporated in common coordinate system directly, see Fig. \ref{fig:CommonCoords+EllipticEgg1}(b).  Other angular coordinate lines are built from  2 pieces of hyperbolas which are smoothly connected to each other on the dashed lines. For this we use additional relations $ah=F(ae)$, e.g. 
\begin{eqnarray}
\;\;ah1_{31}&=f_1\frac{ae_1 cos \gamma_1 +f_1}{ae_1 
+f_1 cos \gamma_1}, \;\;  ah2_{23}=-f_1\frac{ae_1 cos \gamma_2 +f_1}{ae_1 
+f_1 cos \gamma_2};\nonumber \\  
\;\;\overline{ah}3_{31}&=-f_1\frac{\overline{ae}_1 cos \gamma_1 -f_1}{\overline{ae}_1 
-f_1 cos \gamma_1}, \;\; \overline{ah}3_{23}=f_1\frac{\overline{ae}_1 cos \gamma_2 -f_1}{\overline{ae}_1 
-f_1 cos \gamma_2},  
\label{eq6}
\end{eqnarray}
which are the equations of dashed straight lines in the local elliptic coordinates generated by $L_{12}$.  There are 3 local coordinates, so we have  12 equations for our triangle polyelliptic system (other 8 equations are not shown in \eqref{eq6}, but can be written by analogy).
The local angular coordinates $\theta1_{31}$ generated by $L_{12}$ and $\overline{\theta}1_{31}$  generated by left side of $L_{23}$ are related (see  1st eq. in 3rd column of \eqref{eq2}) by the equation on boundary line $1_{31}$, Fig.\ref{fig:EllipticCoords}(b), Fig. \ref{fig:CommonCoords+EllipticEgg1}(a) 
\begin{eqnarray}
f_2 \cos \overline{\theta}1_{31}+f_1\cos \theta1_{31}=f_3 \rightarrow \nonumber \\ \overline{\theta}1_{31}=\arccos[\frac{f_3-f_1 \cos \theta1_{31}}{f_2}], \;\;\; 0\leq \theta1_{31} \leq \gamma_{1}. 
\label{eq7}
\end{eqnarray}
If we assume that origin of $\theta_c$ is $0$ when the local angular coordinate $\theta1_{31}=0$, then the common angular coordinate $\theta_c$  is
\begin{eqnarray}
\theta_c=\theta_{v1}-\overline{\theta}1_{31}, \;\;\; 0\leq \theta_c \leq  \phi_1. 
\label{eq8}
\end{eqnarray}
Here we introduce the angle $\theta_{v1}$ linked to the hyperbola coming through the vertex $1$ of triangle (see Fig. \ref{fig:CommonCoords+EllipticEgg1}(a)): 
\begin{eqnarray}
\theta_{v1}=\arccos[\frac{f_3-f_1}{f_2}], \;\;\;
\phi_1= \theta_{v1} - \gamma_3. 
\label{eq9}
\end{eqnarray}

\begin{figure}
	\centering
\includegraphics[width=0.50\textwidth]{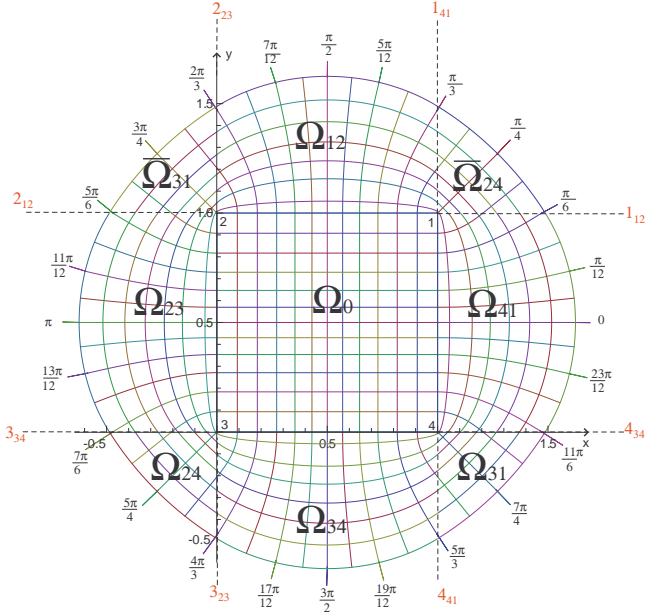}
	\caption{Fig.5. Square 'polyelliptic' coordinates $\mu_c, \theta_c$.  The 'protected area' $\Omega_0$ is covered by Cartesian coordinates with nonuniform scale in $x$- and $y$-directions.}
	\label{fig:Square_WithCentralNet}
\end{figure}

The range $ 0\leq \theta1_{31} \leq \gamma_{1}$  in \eqref{eq7} corresponds to the variation of the angle $\overline{\theta}1_{31}$ inside the part of region  $\overline{\Omega}_{23}$ 
\begin{eqnarray}
\gamma_{3} \leq \overline{\theta}1_{31} \leq \theta_{v1}. 
\label{eq12}
\end{eqnarray}
The range of $\overline{\theta}1_{31} $ is always less than the range of $ \theta1_{31} $ (e.g., compare the data using Fig.\ref{fig:CommonCoords+EllipticEgg1}(a)). This compression reduces the apparent angular span generated by all sides of triangle from $3 \times 180^{\circ}=540^{\circ} $ to the usual $360^{\circ}$, Fig. \ref{fig:CommonCoords+EllipticEgg1}(a). 
The identification of the hyperbola-like coordinates can only be done using the hyperbolas lying in $\overline{\Omega}_{23}, \overline{\Omega}_{31},\overline{\Omega}_{12}$ where they go to infinity.  There is also a simple way to find the coordinate net $\theta_c$ using rectangular plot Fig. \ref{fig:CommonCoords+EllipticEgg1}(b). The horizontal axis presents the $ae_c$ (in logarithmic scale) and vertical axis presents the perimeter or the common angle $\theta_c$. Going through the boundaries of 6 regions and  6 additional adjacent regions with hyperbola-like coordinates, we can receive the coefficients $A_i, B_i$ in  \eqref{eq1} for 12  regions (for short we use here the radial coordinate $ae_c=(f_1+f_3)\cosh \mu_c)$):
\begin{eqnarray}
\begin{array}{ll}
\overline{A}_2=ae_c\cos[\theta -\theta_{v1}], \; 
\overline{B}_2=\sqrt{(ae_c)^2-f_2^2} \sin[\theta_c -\theta_{v1}];  \\
\overline{A}_{2F}=(ae_c-f_3)\left(\frac{f_3-f_2 \cos[\theta -\theta_{v1}]}{f_1}\right),  \\
\overline{B}_{2F}=\sqrt{\left((ae_c-f_3)^2-f_1^2\right)\left(1-\left(\frac{f_3-f_2 \cos[\theta_c -\theta_{v1}]}{f_1}\right)^2\right)}; \\
A_{1}=(ae_c-f_3)\cos\left[\theta_c -\left(\theta_{v1}-\gamma _3-\gamma _1\right)\right],  \\
B_{1}=\sqrt{(ae_c-f_3)^2-f_1^2}\sin\left[\theta_c -\left(\theta_{v1}-\gamma _3-\gamma _1\right)\right]; \\
\overline{A}_{3B}=(ae_c-f_3)\left(\frac{-f_2-f_3 \cos\left[\theta_{v1}+\gamma _1+\gamma _2-\theta_c \right]}{f_1}\right),  \\
\overline{B}_{3B}=\sqrt{(ae_c-f_3)^2-f_1^2}\times  \\
\sqrt{1-\left(\frac{-f_2-f_3 \cos\left[\theta_{v1}+\gamma _1+\gamma _2-\theta_c \right]}{f_1}\right){}^2}; \\
\overline{A}_3=(ae_c-f_3+f_2)\left(\cos\left[\theta_{v1}+\gamma _1+\gamma _2-\theta_c \right]\right),  \\
\overline{B}_3=\sqrt{(ae_c-f_3+f_2)^2-f_3^2}\sin\left[\theta_{v1}+\gamma _1+\gamma _2-\theta_c \right]; \\
\overline{A}_{3F}=(ae_c-f_3+f_2-f_1)\left(\frac{f_1-f_3 \cos\left[\theta_{v1}+\gamma _1+\gamma _2-\theta_c \right]}{f_2}\right),  \\
\overline{B}_{3F}=\sqrt{(ae_c-f_3+f_2-f_1)^2-f_2^2}\times  \\
\sqrt{1-\left(\frac{f_1-f_3 \cos\left[\theta_{v1}+\gamma _1+\gamma _2-\theta_c \right]}{f_2}\right){}^2}; \\
A_2=(ae_c-f_3+f_2-f_1)(\cos[\theta_c -\theta_{v1}]),  \\
B_2=\sqrt{(ae_c-f_3+f_2-f_1)^2-f_2^2}\sin[\theta_c -\theta_{v1}]; \\
\overline{A}_{1B}=(ae_c-f_3+f_2-f_1)\frac{-f_3-f_1 \cos\left[\theta_{v1}+\gamma _1+2\gamma _2+\gamma _3-\theta_c \right]}{f_2},  \\
\overline{B}_{1B}=\sqrt{(ae_c-f_3+f_2-f_1)^2-f_2^2}\times  \\
\sqrt{1-\left(\frac{-f_3-f_1 \cos\left[\theta_{v1}+\gamma _1+2\gamma _2+\gamma _3-\theta_c \right]}{f_2}\right){}^2}; \\
\overline{A}_1=(ae_c+f_2-f_1)\left(\cos\left[\theta_{v1}+\gamma _1+2\gamma _2+\gamma _3-\theta_c \right]\right),  \\
\overline{B}_1=\sqrt{(ae_c+f_2-f_1)^2-f_1^2}\times  \\
\sin\left[\theta_{v1}+\gamma _1+2\gamma _2+\gamma _3-\theta_c \right]; \\
\overline{A}_{3F}=(ae_c-f_1)\left(\frac{f_2-f_1 \cos\left[\theta_{v1}+\gamma _1+2\gamma _2+\gamma _3-\theta_c \right]}{f_3}\right),  \\
\overline{B}_{3F}=\sqrt{(ae_c-f_1)^2-f_3^2}\times  \\
\sqrt{1-\left(\frac{f_2-f_1 \cos\left[\theta_{v1}+\gamma _1+2\gamma _2+\gamma _3-\theta_c \right]}{f_3}\right){}^2}; \\
A_3=(ae_c-f_1)\left(\cos\left[\theta_c -\left(\theta_{v1}+\pi  -\gamma _3\right)\right]\right),  \\
B_3=\sqrt{(ae_c-f_1)^2-f_3^2}\sin\left[\theta_c -\left(\theta_{v1}+\pi  -\gamma _3\right)\right]; \\
\overline{A}_{2B}=(ae_c-f_1)\left(\frac{-f_1-f_2 \cos[\theta_{v1}+2\pi -\theta_c ]}{f_3}\right),  \\
\overline{B}_{2B}=\sqrt{(ae_c-f_1)^2-f_3^2}\times  \\
\sqrt{1-\left(\frac{-f_1-f_2 \cos[\theta_{v1}+2\pi -\theta_c ]}{f_3}\right)^2}.\end{array} \label{eq13}
\end{eqnarray}
These long but rather simple expressions allow to build any  polyelliptic coordinates outside the arbitrary triangle, Fig.\ref{fig:TriangleCoords_Small-Large}.  If we are moving around counterclockwise, the coefficients with index ``B'' correspond to compressed region before of $\overline{\Omega}$, coefficients with index ``F'' correspond to compressed region in front of $\overline{\Omega}$.  As example, the shaded area in  Fig.\ref{fig:TriangleCoords_Small-Large}(a) includes 3 regions: with indices ``B'',''F'' and  region $\overline{\Omega}_{31}$ itself.    
If we take degenerated triangle $f_1=f_2$, $f_3=0$, the triangle polyelliptic system becomes the simple elliptic coordinates.  Taking $f_1=f_2=f_3=f$, we can produce the equilateral triangle polyelliptic system, Fig. \ref{fig:EquilateralEllipticTriangle_Small}.  It is not difficult to modify the equations \eqref{eq13} and build also the square polyelliptic system, Fig.\ref{fig:Square_WithCentralNet}. The interesting feature of the square and rectangular polyelliptic system is the absence of the regions with true hyperbolas. All angular coordinates (except 8 straight lines coming out from vertices and centers of interfocal lines) are built from two pieces of hyperbolas.
The construction of n-polygon polyelliptic system may include the angular coordinates which are built from $n-1$ pieces of hyperbolas.

\begin{figure}
	\centering
		\includegraphics[width=0.50\textwidth]{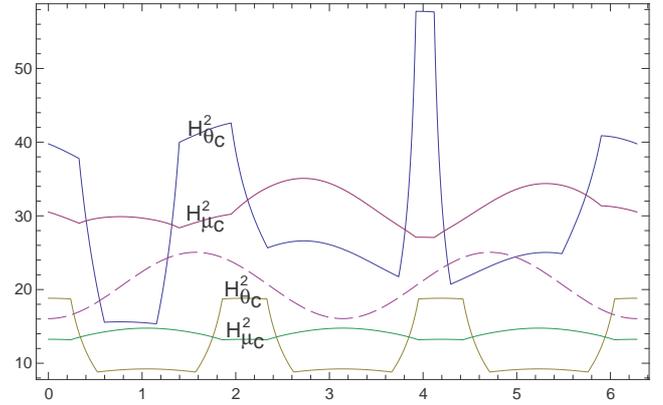}
	\caption{Fig.6. Scale factors: $H_{\theta_c}^2$,  $H_{\mu_c}^2$  in  triangle coordinates Fig.\ref{fig:TriangleCoords_Small-Large} (blue, red), Fig.\ref{fig:EquilateralEllipticTriangle_Small} (yellow, green),  calculated for fixed $\mu_c=1.1$. Dashed curve presents Mathieu's elliptic scale factors for $f=3$, $\mu=1.1$.}
	\label{fig:LameCoeff}
\end{figure}

\begin{figure}
	\centering		
	\includegraphics[width=0.50\textwidth]{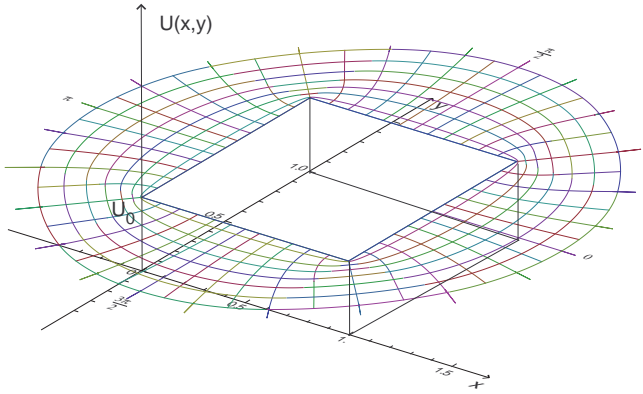}
	\caption{Fig.7. Square-well potential with square shape. In 'square-elliptic' coordinates the problem of finding the energy levels and wave eigenfunctions is solvable. }
	\label{fig:SquareWell_SquareShape_3a}
\end{figure}

\textbf{II. St\"{a}ckel form and differential equations.} 
The HE $\Delta \Psi+k^2\Psi=0$ in 2D Cartesian coordinates can be written in a general curvilinear orthogonal form
\begin{eqnarray}
\frac{1}{H_1H_2}\left[\partial_1\left(\frac{H_2}{H_1}\partial_1 \Psi\right)+\partial_2\left(\frac{H_1}{H_2}\partial_2 \Psi\right)\right]+k^2\Psi=0,
\label{eqg1}
\end{eqnarray}
where $H_i=\sqrt{g_{i,i}}$ are scale factors, $\partial_i=\partial /\partial q_i$. 
We can show that the metric $g_{i,j}$ for polyelliptic coordinates $\theta_c,\mu_c$  in the exterior of n-poligon are diagonal and $H_{\theta_c}=\sqrt{(\partial_{\theta_c} x)^2+(\partial_{\theta_c} y)^2}$, $H_{\mu_c}=\sqrt{(\partial_{\mu_c} x)^2+(\partial_{\mu_c} y)^2}$ are continuous functions of $\theta_c, \mu_c$ having the St\"{a}ckel expression \cite{Stackel:1890}  
\begin{eqnarray}
H_{\theta_c}^{2}=g_1(\theta_c)[h_1(\mu_c)+h_2(\theta_c)],\nonumber \\ H_{\mu_c}^{2}=g_2(\mu_c)[h_1(\mu_c)+h_2(\theta_c)],
\label{eq16}
\end{eqnarray}
where $g_1,g_2,h_1,h_2$ are continuous functions consisting of pieces from trigonometric or hyperbolic functions. It is clear that derivatives of \eqref{eq16} will be  piecewise continuous functions having a finite jumps on the boundaries of the regions. As exsample in the region $\overline{\Omega}_{23}$ ($\overline{A}_2, \; \overline{B}_2$ are taken from \eqref{eq13}) these scale factors are 
\begin{eqnarray}
H_{\theta_c}^{2}=(f_1+f_3)^2 \cosh[\mu_c]^2 -f_2^2\cos[\theta_c-\theta_{v1}]^2, 
\label{eq17a} \\
H_{\mu_c}^{2}=\frac{(f_1+f_3)^2}{(f_1 + f_3)^2 \cosh[\mu]^2 -f_2^2 }\times \nonumber \\
((f_1+f_3)^2 \cosh[\mu_c]^2 -f_2^2\cos[\theta_c-\theta_{v1}]^2), 
\label{eq17b}
\end{eqnarray}
where $g_1(\theta_c)=1,$ and $ g_2(\mu_c)$ is 1st term of \eqref{eq17b}.
For short, we skip the similar expression for other regions. The Fig.\ref{fig:LameCoeff} illustrates these scale factors as a function of $\theta_c$ for arbitrary and equilateral triangle coordinates, Fig.\ref{fig:EquilateralEllipticTriangle_Small}, as well as Mathieu's scale factors for comparison. Searching for solutions in the form $\Psi=\Psi_1(\theta_c)\Psi_2(\mu_c)$, the HE \eqref{eq16} can be split up into two ordinary differential equations with separation constant $\lambda$:
\begin{eqnarray}
\partial^2_{\theta_c}\Psi_1 -\frac{g_1^{'}}{2g_1}\partial_{\theta_c}\Psi_1 +g_1[\lambda + k^2 h_2]\Psi_1=0,\label{eq18a} \\ \partial^2_{\mu_c}\Psi_2 -\frac{g_2^{'}}{2g_2}\partial_{\mu_c}\Psi_2-g_2[\lambda -  k^2 h_1]\Psi_2=0.
\label{eq18b}
\end{eqnarray}
If the ODE \eqref{eq18a} can be solved in the exterior of polygon with periodic condition $\Psi_1(0)=\Psi_1(2\pi)$, it gives a new set of eigenfunctions specific for this particular polygon with the corresponding eigenvalues. The equation \eqref{eq18b} should give the radial solution for HE. Note that set ODE \eqref{eq18a}-\eqref{eq18b} becomes  Mathieu's equations\cite{MorseFeshbach:1953} when we use the elliptic scale factors  $H_{\theta_c}=H_{\mu_c}=\sqrt{\cosh^2_{\mu_c}-\cos^2_{\theta_c}}$, where $g_1$, $g_2$ are constants.     

\textbf{III. Applications.} There are many possible applications of polyelliptic coordinates, and we outline only several.

(a) The n-polygon polyelliptic system has a 'protected' area, which can be thought of as a infinite high potential. If a plane wave is expanded on eigenfunctions of ODE  \eqref{eq18a}-\eqref{eq18b}  with Dirichlet condition on perimeter of polygon, we should immediately receive exact solution for scattering problem in variety of applications which employ the sharp edges (acoustics, radar, etc.).

(b) The rotation of equilateral Fig.\ref{fig:EquilateralEllipticTriangle_Small} or isosceles triangle elliptic coordinates around axis of symmetry creates the new types of 3D polyelliptic coordinates which are neither prolate or oblate. The 'protected' volume will be a cone and finding the eigenfunctions can give the solution for this important 3D scattering problem.    
  
(c) SE for 2D potential,  Fig. \ref{fig:SquareWell_SquareShape_3a},
\begin{eqnarray}
U(x,y) = \left \{\begin{array}{ll} 0,&  \;  0 \leq x\leq a,\; 0 \leq y \leq a;\\  
U_0,&  \;  x<0, x > a, \;y<0, y>a,
\end{array} \right.
\label{eqc20}
\end{eqnarray}
as previously thought, does not have a separable system\footnote{HE for 2D box with infinite high potential wall was solved by Poisson \cite{Poisson:1828}.  The problem of scattering on square box (finite of infinite) and spectral problem for potential \eqref{eqc20} have not been exactly solved yet. But due to importance of this problem in fiber optics, there are several numerical solutions\cite{Goell:1969},\cite{Marcatili:1969}} because the square perimeter does not correspond to any coordinate line. However, it does correspond to $\mu_c=0$ in square polyelliptic system Fig.\ref{fig:Square_WithCentralNet},\ref{fig:SquareWell_SquareShape_3a} and the surrounding of square has two independent wave functions $\Psi_1, \Psi_2$ which allow to solve the spertrum problem with potential  \eqref{eqc20} if eigenfunctions of \eqref{eq18a}-\eqref{eq18b} are found.

\textbf{IV. Summary.} We have presented an infinite family of new orthogonal coordinate systems which admit the separation of variables for SE and HE in 2D space and assume the existence of such system in 3D space. Each coordinate system is related to one particular polygon and has two ODE describing a set of eigenfunctions with corresponding eigenvalues. If the eigenfunctions outside the polygon can be calculated, this approach may  help to solve some long standing problems.

%\bibliographystyle{apsrev}
%\bibliography{../../Focusing_and_Channeling_in_Crystals/chan02}

\begin{thebibliography}{99}
%
%
\bibitem{RobEis}
 H.\,P. Robertson, Math. Ann.  {\bf 98}, 749 (1928); L.\,P. Eisenhart, Ann. Math. {\bf 35}, 284, (1934); Phys. Rev. {\bf 45}, 427 (1934); {\bf 74}, 87 (1948).
%

%
\bibitem{MoonSpencer:1988}
P. Moon and D.\,E. Spencer,  Field Theory Handbook, Springer-Verlag, Berlin, 1988.
%

\bibitem{MorseFeshbach:1953}
P.\,M. Morse  and H. Feshbach,  Methods of Theoretical Physics, vol.1-2, McGraw-Hill Book Co., New York, 1953. 

\bibitem{Miller:1976}
W. Miller, Jr. Symmetry and separation of variable, Addison 
Wesley, 1977.
%


\bibitem{Kalins:1986}
E.\,G. Kalnins, Separation of Variables for Riemannian Spaces of Constant Curvature, Longman Scientific and Technical, Essex, 1986.
%

%
\bibitem{Fedoryuk:1989}
 M.\,V. Fedoryuk,  Math. Notes {\bf 46}, 804-811 (1989). 
%

\bibitem{PDG:2004}
V. Rokhlin and Hong Xiao,  Appl. Comput. Harmon. Anal. {\bf 22}, 105-123 (2007).
%

\bibitem{Slepian:1983}
 D. Slepian, SIAMReview, {\bf 25}, Issue 3,  379-393 (1983). 

\bibitem{Stackel:1890} 
 P. Stackel, Mathematische Annalen, {\bf 35}, 91-103 (1890). 


\bibitem{Poisson:1828} 
 S.D. Poisson, ``Memoire sur l'equilibre et le mouvement des corps 
elastiques'' in ``Memoires de l'Institut'', v.VIII Paris, 1828. 


\bibitem{Goell:1969} 
 J.E. Goell, Bell. Syst. Tech. J., {\bf 48}, 2133 (1969). 

\bibitem{Marcatili:1969} 
E.A.J. Marcatili, Bell. Syst. Tech. J., {\bf 48}, 2071 (1969).






\end{thebibliography}

%\begin{thebibliography}{99}
%\bibitem{kulik}
%I.\,O. Kulik and A.\,N. Omel'yanchuk, Fiz.~Nizk.~Temp. {\bf 3}, 945
%(1977) [translated in Sov. J.~Low~Temp.~Phys.].

%\bibitem{been1}
%C.\,W. Beenakker and H. van Houten, Phys.~Rev.~Lett. {\bf 66}, 3056
%(1991).

%\bibitem{shumeiko1}
%V.\,S. Shumeiko, E.\,N. Bratus', G. Wendin, Fiz.~Nizk.~Temp. {\bf 23},
%249 (1997) [Sov.~J.~Low Temp.~Phys. {\bf 23}, 181 (1997)],
%cond\D mat/9610101.

%\end{thebibliography}

\end{document}